\documentclass[english,superscriptaddress]{revtex4}
\usepackage[T1]{fontenc}
\usepackage[latin1]{inputenc}
\usepackage{graphicx}

\makeatletter

\usepackage{babel}
\makeatother
\begin{document}

\title{A study of elliptic flows in a quark combination model}

\author{Tao Yao}

\affiliation{Department of Physics, Shandong University, Jinan, Shandong 250100,
P. R. China}

\author{Qu-bing Xie}

\affiliation{Department of Physics, Shandong University, Jinan, Shandong 250100,
P. R. China}

\author{Feng-lan Shao}

\affiliation{Department of Physics, Qufu Normal University, Qufu, Shandong 273165,
P. R. China}

\begin{abstract}
We carry out a detail study of elliptic flows in Au-Au collisions
at 200 AGeV in a quark combination model. We find that elliptic flow
data for a variety of hadrons can be well reproduced except pions
if constituent quarks with equal parallel transverse momenta combine
into initially produced hadrons. In a combination mechanism where
initial hadrons are formed by quarks with unequal parallel transverse
momenta, theoretical predictions agree with data for all available
hadrons including pions. The mass hierarchy at low transverse momenta
in elliptic flows can be understood in the same quark combination
mechanism as in the mediate range of transverse momenta.
\end{abstract}
\maketitle \noindent \hspace{1cm} {\bf PACS number:} 25.75.-q,
25.75.Ld, 12.40.-y

\noindent \hspace{1cm} {\bf Keywords:} Heavy ion collision,
collective flow, quark matter, quark combination model, the mass
hierarchy of elliptic flows

\section{Introduction}

The elliptic flow is one of the most important observables in
ultra-relativistic heavy ion collisions, which carries a lot of
information about the initial state of the dense matter
\cite{Ollitrault:1992bk,Sorge:1998mk,Barrette:1994xr,Appelshauser:1997dg}.
The elliptic flow $v_{2}$ for a hadron is defined as the coefficient
of the term $\cos2\phi$ in the Fourier expansion of the invariant
differential cross section $E\frac{d^{3}N}{d^{3}P}$, where $\phi$ is
the azimuthal angle of the hadron momentum in the transverse plane
\cite{Voloshin:1994mz,Poskanzer:1998yz}. The E877 collaboration
first measured the anisotropic flow in heavy ion collisions at AGS
energies. In recent years, a thorough experimental study of elliptic
flows for various hadrons has been taken at relativistic heavy ion
collider (RHIC) energies. The elliptic flows as functions of
centrality, pseudorapidity and transverse momentum have been
measured. Various theoretical models have been proposed to describe
data, such as coalescence or recombination models
\cite{Molnar:2003ff,Hwa:2002tu,Fries:2003vb,Greco:2003xt}, transport
models \cite{Bleicher:2000sx,Lin:2001zk,Teaney:2001gc}, and
hydrodynamic models
\cite{Kolb:2000sd,Huovinen:2001cy,Borghini:2005kd} etc..

A salient feature of $v_{2}(p_{T})$ at mediate transverse momenta
$p_{T}$ is the quark number scaling, i.e. when both $v_{2}$ and
$p_{T}$ rescaled by the constituent quark number $n_{q}$ of the
hadron, the elliptic flow is a universal function of $p_{T}$ for all
hadrons. The quark number scaling of the elliptic flow and the
abnormal ratio of baryons to mesons at mediate $p_{T}$ support the
quark recombination mechanism in hadron formation. However, at small
transverse momenta $p_{T}$ the quark number scaling is not obvious,
instead, there is a different structure the so-called mass hierarchy
\cite{Adams:2004bi} in data. In hydrodynamic models, the mass
hierarchy arises from radial flows which are hadron masses
\cite{Huovinen:2001cy} dependent. But hydrodynamic models cannot
describe elliptic flow data at mediate and high $p_{T}$. The mass
hierarchy structure of $v_{2}(p_{T})$ for charged pions and kaons
and protons/anti-protons has been reproduced by a multi-phase
transport model \cite{Lin:2004en}. In the large $p_{T}$ range of
about 10 GeV/c, Ref. \cite{Nonaka:2003hx,Greco:2003mm,Fries:2003kq}
have described the global behavior of $v_{2}(p_{T})$ for some
hadrons in combination mechanism, but the mass hierarchy of
$v_{2}(p_{T})$ for various hadrons at low $p_{T}$ has not been fully
compared with experiment data or reproduced within
coalescence/recombination models. Ref.
\cite{Molnar:2004rr,Pratt:2004zq} and some models have given several
possible origins of the mass hierarchy, however, we show decay is
one of the most important reasons and give another explanation for
the mass hierarchy. In this paper, we carry out a detail study of
elliptic flows in a quark combination model in the whole $p_{T}$
range with a special focus on the mass hierarchy or fine structure
at low $p_{T}$. We will show that the fine structure of elliptic
flows can be well described in the quark combination picture
provided initially produced hadrons and resonance decays are treated
separately and properly.

\section{Quark combination model}

All kinds of hadronization models demand themselves, consciously or
not, satisfy rapidity or momentum correlation for quarks in the neighborhood
of phase space. The essence of this correlation is its agreement with
the fundamental requirement of QCD which uniquely determines the quark
combination rule \cite{Xie:1988wi,Xie:1997ap}. According to QCD,
a $q\overline{q}$ may be in a color octet or a singlet which means
a repulsive or an attractive interaction between them. The smaller
the difference in rapidity for two quarks is, the longer the interaction
time is. So there is enough time for a $q\overline{q}$ to be in a
color singlet and form a meson. Similarly, a $qq$ can be in a sextet
or an anti-triplet. If its nearest neighbor is a $q$ in rapidity,
they form a baryon. If the neighbor is a $\overline{q}$, because
the attraction strength of the singlet is two times that of the anti-triplet,
$q\overline{q}$ will win the competition to form a meson and leave
a $q$ alone.

We have developed a variant of the quark combination model (QCM) based
on a simple quark combination rule \cite{Xie:1988wi,Xie:1997ap} incorporating
the above QCD requirements. The flavor SU(3) symmetry with strangeness
suppression in the yields of initially produced hadrons is fulfilled
in our QCM \cite{Xie:1988wi,Wang:1995ch}. Using our QCM, we have
described most of multiplicity data for hadrons in electron-positron
and proton-proton/anti-proton collisions \cite{Xie:1988wi,Liang:1991ya,Wang:1995ch,Zhao:1995hq,Wang:1996jy,Si:1997ux}.
Also we solved a difficulty facing other QCMs in describing the TASSO
data for baryon-antibaryon correlation in electron-positron collisions:
they can be successfully explained by our QCM \cite{Si:1997ux,Xie:1997ap}.
Our QCM can also combine with the color flow picture \cite{Wang:1996pg}
to describe the hadroniztion of multiparton states \cite{Wang:1995gx,Wang:1999xz,Wang:2000bv}.
We have extended our QCM to reproduce the recent RHIC data for hadron
multiplicities and $p_{T}$ spectra \cite{Shao:2004cn}.

In this paper we present a detail analysis of $v_{2}(p_{T})$ in the
whole $p_{T}$ range based on our QCM. We put particular emphasis on
the low $p_{T}$ region where the elliptic flow data for various
hadrons show non-trivial fine structure. Our treatment of elliptic
flows is different from other coalescence or recombination models
\cite{Nonaka:2003hx,Greco:2003mm}. We extract $v_{2,q}(p_{T})$ for
quarks as input from fitting the experiment data of various hadrons.
So we can construct elliptic flows for all initially produced
hadrons from $v_{2,q}(p_{T})$ for quarks in our QCM. Then we let all
resonances decay by using the decay program of the event generator
PYTHIA \cite{Sjostrand:2003wg} to obtain elliptic flows for final
state hadrons. One advantage of using PYTHIA to deal with decays is
that all available decay channels are covered. Here we only take
into account light quarks in hadronization. In central Au-Au
collisions at 200 AGeV, the yield of charm hadrons is about 28 in
the whole rapidity range \cite{Adler:2004ta}. So charm quarks are
negligible compared to light quarks in $v_{2}(p_{T})$ for final
state hadrons.

\begin{figure}

\caption{\label{cap:fig1}The $v_{2}(p_{T})$ spectra for final state anti-proton
from Eq. (\ref{eq:v2q0}) (dashed line) and (\ref{eq:v2q}) (solid
line). The sub-diagram in the upper-left corner shows curves of Eq.
(\ref{eq:v2q0}) (dashed line) and (\ref{eq:v2q}) (solid line). }

\includegraphics[%
  scale=0.4]{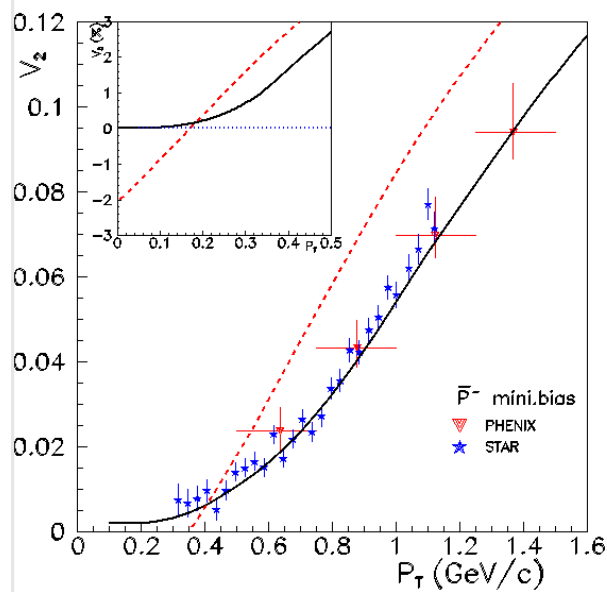}
\end{figure}

\begin{figure}

\caption{\label{cap:fig2}The $v_{2}(p_{T})$ spectra for directly produced
(upper panel) and final state (lower panel) hadrons from Eq. (\ref{eq:v2q}). }

\includegraphics[%
  scale=0.6]{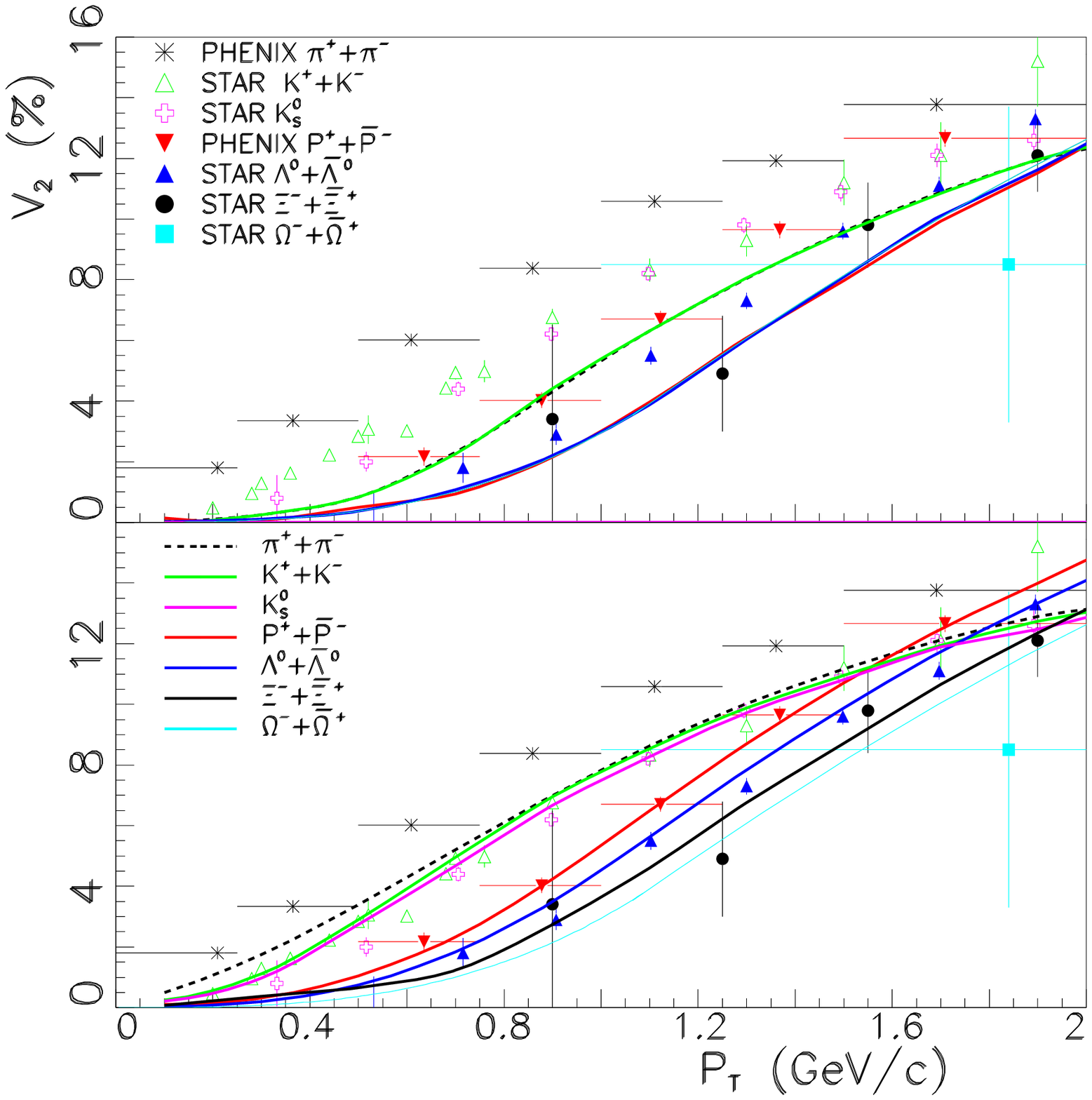}
\end{figure}

\begin{figure}

\caption{\label{cap:fig3}The $v_{2}(p_{T})$ spectra for directly produced
(upper panel) and final state (lower panel) hadrons from Eq. (\ref{eq:v2qi}). }

\includegraphics[%
  scale=0.6]{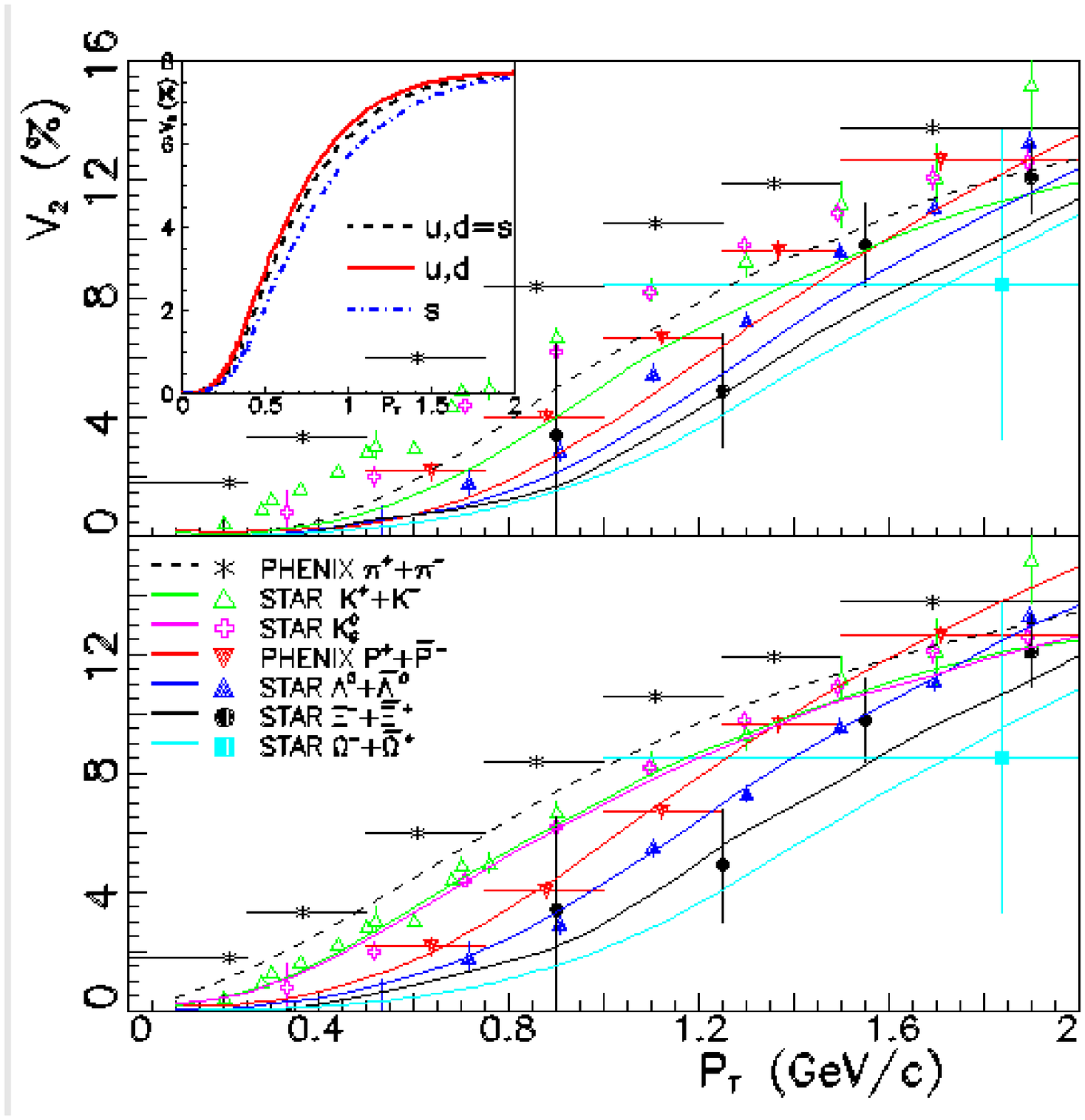}
\end{figure}

\begin{figure}

\caption{\label{cap:fig4}The $v_{2}(p_{T})$ spectra for final state hadrons
are given in the upper panel. The $v_{2}(p_{T})$ spectra scaled by
constituent quark number are given in the lower panel. The theoretical
curves are results in the unequal momentum combination scheme. }

\includegraphics[%
  scale=0.6]{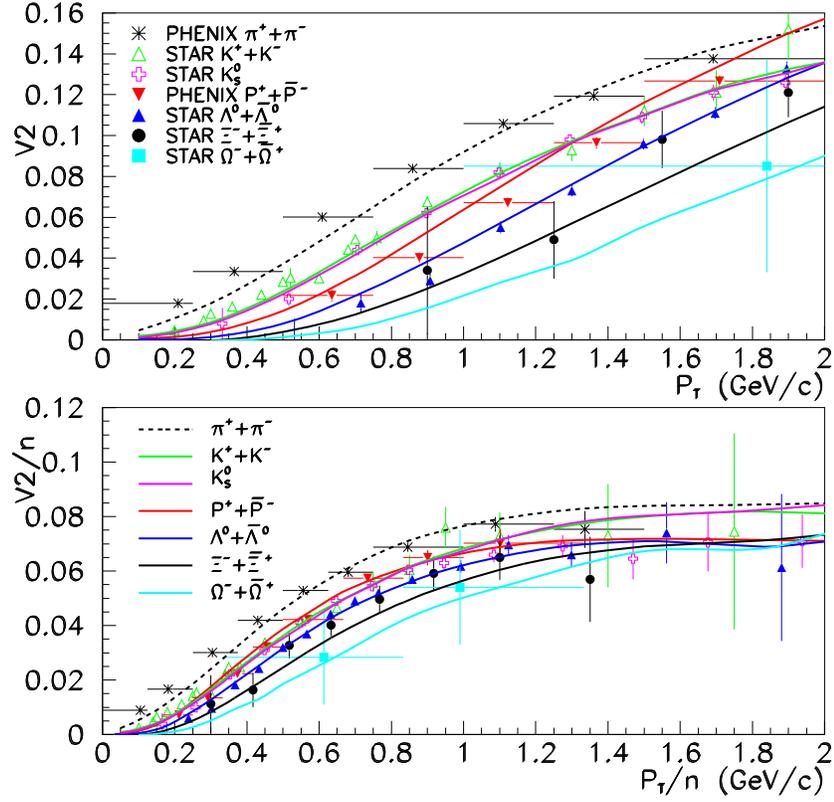}
\end{figure}

\begin{figure}

\caption{\label{cap:fig5}The $v_{2}(p_{T})$ spectra for $\phi$ mesons in
three combination schemes. }

\includegraphics[%
  scale=0.4]{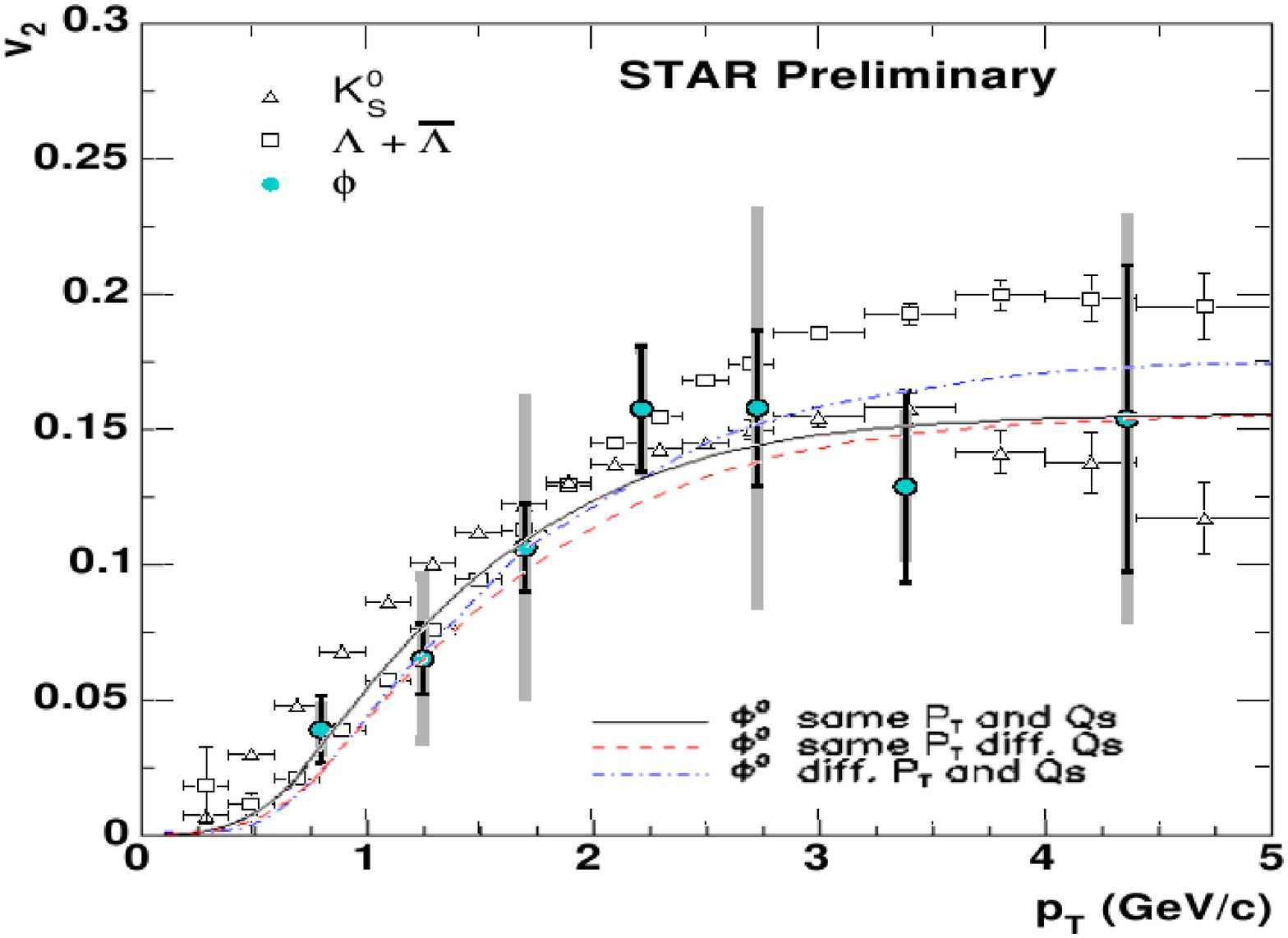}
\end{figure}

\section{Results and analysis}

In order to describe data of elliptic flows, only those quarks with
collinear momenta are allowed to combine into hadrons, because the
combination of constituent quarks in different momentum directions
would lead to large invariant masses which are off-mass-shell for
most hadrons. To describe data in central rapidity, we choose the
rapidity range $y\in[-0.5,0.5]$ for hadrons in our calculation. We
also assume that quarks and antiquarks of a same flavor have the the
same $v_{2,q}(p_{T})$. Considering that the coalescence or recombination
models work very well at mediate $p_{T}$, we focus our study on elliptic
flows at low $p_{T}$ which have not been well investigated in combination-like
models. We assume three collinear combination strategies, the naive
scheme with equal $p_{T}$ combination and the same $v_{2,q}(p_{T})$
for quarks of all flavors, the modified naive scheme with equal $p_{T}$
combination and a different $v_{2,q}(p_{T})$ for strange quarks from
light quarks, and the scheme with different $p_{T}$ combination.

\subsection{Naive combination scheme}

In this scheme, all $u$, $d$ and $s$ quarks are assumed to have
the same $v_{2q}(p_{T})$ spectrum. The quarks with equal $p_{T}$
combine into hadrons. Our input is based on \begin{equation}
v_{2q}^{(0)}(p_{T})=0.078\tanh(1.59p_{T}-0.27)\label{eq:v2q0}\end{equation}
 for quarks given in Ref. \cite{Greco:2004ex} by fitting the proton
data. Hereafter the unit of $p_{T}$ is GeV. We make two
improvements, one is that the decay contributions have been
subtracted, the other is that the negative part of
$v_{2q}^{(0)}(p_{T})$ inconsistent to data and to the expected
quadratic behavior \cite{Danielewicz:1994nb,Molnar:2004rr} at small
$p_{T}$ (see the dashed line in Fig. 1) is corrected to be positive.
The improved $v_{2,q}(p_{T})$ is given by
\begin{eqnarray}
v_{2,q}(p_{T}) & = & 0.256p_{T}^{3},\;\;\;0\leq p_{T}\leq0.365,\nonumber \\
v_{2,q}(p_{T}) & = & 0.078\tanh[1.45(p_{T}-0.25)],\;\;\;
p_{T}>0.365.\label{eq:v2q}\end{eqnarray} Note that considering the
continuity of the Eq. (\ref{eq:v2q}), we adopt the $p_{T}^{3}$
instead of the $p_{T}^{2}$ form. The difference resulted from the
two function forms at low $p_{T}$ range can be neglected for the
coefficient of $p_{T}^{3}$ or $p_{T}^{2}$ is adjustable. One sees
that the data for antiprotons are well reproduced at small $p_{T}$
with the above input, see the solid line in Fig. \ref{cap:fig1}.
Moreover it can fit all hadron data except pions as shown in the
lower panel of Fig. \ref{cap:fig2}. Note that in figures following
Fig. \ref{cap:fig2} the order of curves from left to right is the
same that of data for various hadrons. The STAR data in Fig.
\ref{cap:fig1} are taken from Ref. \cite{Adams:2004bi}. In Fig.
\ref{cap:fig1} and \ref{cap:fig2}, the PHENIX data are taken from
Ref. \cite{Adler:2003kt}. In Fig. \ref{cap:fig2}, the data for
$K_{S}^{0}$ and $\Lambda+\overline{\Lambda}$ are from Ref.
\cite{Adams:2003am}, the data for $\Xi^{-}+\overline{\Xi}^{+}$ and
$\Omega^{-}+\overline{\Omega}^{+}$ are from Ref.
\cite{Adams:2005zg}, and the data for kaons are from Ref.
\cite{Adams:2004bi}.

In the naive combination scheme, directly produced mesons and baryons
have the same elliptic flows $v_{2,M}(p_{T})=2v_{2,q}(p_{T}/2)$ and
$v_{2,B}(p_{T})=3v_{2,q}(p_{T}/3)$ respectively, regardless of their
masses and species. Comparing the upper and lower panels of Fig. \ref{cap:fig2},
one observes that the $v_{2}(p_{T})$ curve for the hadron $i$ shifts
somewhere in the left depending on how much the decay contributions
the hadron has. One sees that the fine structure of $v_{2}(p_{T})$
for hadrons at small $p_{T}$ results from decay contributions. But
for pions it is not enough to reproduce the data by only including
the decays, one has to take into account other effects. We will address
this issue in the modified naive combination scheme in the next subsection.

\subsection{Modified naive combination scheme}

In this scheme, we take into account that $v_{2,q}(p_{T})$ for strange
quarks is different from light quarks. The difference is mainly at
small $p_{T}$ since the quark mass effects barely influence the magnitude
of $v_{2}(p_{T})$ at mediate $p_{T}$, see the attached figure in
the upper-left corner in the upper panel of Fig. \ref{cap:fig3}.
Note that Ref. \cite{Nonaka:2003hx,Greco:2003mm} also include the
strange-light difference. We know that the strongly coupled quark-gluon-plasma
produced at RHIC is relativistic fluid, then we can assume that the
average transverse momenta for $u$, $d$ and $s$ quarks are proportional
to their average energies, $\frac{<p_{T}>_{u,d}}{<p_{T}>_{s}}\approx\frac{<E>_{u,d}}{<E>_{s}}$.
Based on this relation, we modify $v_{2,q}(p_{T})$ for $u$, $d$
and $s$ quarks as, \begin{eqnarray}
v_{2,i}(p_{T}) & = & 0.256(p_{T}/\alpha_{i})^{3},\;\;\;0\leq p_{T}/\alpha_{i}\leq0.365,\nonumber \\
v_{2,i}(p_{T}) & = & 0.078\tanh[1.45(p_{T}/\alpha_{i}-0.25)],\;\;\; p_{T}/\alpha_{i}>0.365.\label{eq:v2qi}\end{eqnarray}
where $i=u,d,s$ and $\alpha_{i}$ is given by \begin{eqnarray}
\alpha_{u,d} & = & \frac{2<E>_{u,d}}{<E>_{u,d}+<E>_{s}}\approx0.94,\nonumber \\
\alpha_{s} & = &
\frac{2<E>_{s}}{<E>_{u,d}+<E>_{s}}\approx1.12.\end{eqnarray} where
the average values of quark energies can be approximated as
$\left\langle E\right\rangle _{f}\approx\sqrt{m_{f}^{2}+\left\langle
p_{T}^{2}\right\rangle }$ with constituent quark mass $m_{f}$ for
the flavor $f$ ($m_{u,d}=0.34$ GeV, $m_{s}=0.5$ GeV) and with
$\left\langle p_{T}^{2}\right\rangle $ given by the distribution
$f(p_{T})=(p_{T}^{2.3}+p_{T}^{0.2}+1)^{-3.0}$ from Ref.
\cite{Shao:2004cn}. Here we have implied the central rapidity
$y\approx0$ in estimating the average energy. Based on the input for
quark elliptic flows in Eq. (\ref{eq:v2qi}), $v_{2}(p_{T})$ for
meson and baryon are written as \begin{eqnarray}
v_{2,M}(p_{T}) & \approx & v_{2,i}(p_{T}/2)+v_{2,j}(p_{T}/2),\nonumber \\
v_{2,B}(p_{T}) & \approx & v_{2,i}(p_{T}/3)+v_{2,j}(p_{T}/3)+v_{2,k}(p_{T}/3),\end{eqnarray}
with $i,j$ denote the constituent quarks in the meson $M$ and $i,j,k$
denote those in the baryon $B$. Therefore the $v_{2}(p_{T})$ curves
for hadrons with the same flavor content are identical. The results
in this combination scheme are shown in Fig. \ref{cap:fig3}. In contrast
to Fig. \ref{cap:fig2}, the $v_{2}(p_{T})$ curves for directly produced
hadrons split according to their flavor contents and lead to even
larger splittings for final state hadrons. The pion curve is partially
improved but not enough to match the data. For other hadrons, the
agreement between theoretical predictions and data is satisfactory.
Although tuning $\alpha_{u,d}$ to a smaller value would make pion
curve fit the data, the fine structure of $v_{2}(p_{T})$ for other
hadrons prevent them from splitting too much. The results show that
in the quark combination mechanism the quark mass effect and resonance
decays partially account for the fine structure of $v_{2}(p_{T})$
for final state hadrons. In contrast to the hydro-model interpretation
of differences among hadron $v_{2}(p_{T})$ as hadron mass effects,
the mass effects at the quark level is partially responsible for such
differences in the quark combination picture. To identify the mass
effects of strange quarks and light quarks, a better way is to measure
and compare $v_{2}(p_{T})$ for $\phi$ and $K^{*}$ in experiments.

\subsection{Unequal $p_{T}$ combination scheme}

As shown in Fig. \ref{cap:fig3}, it is still not enough to reproduce
the pion data by including resonance decays and the quark mass
effect. In order to better fit the pion data together with all other
hadrons, we relax in this scheme the constraint that quarks with
equal $p_{T}$ combine into hadrons by allowing unequal $p_{T}$
combination. The unequal $p_{T}$ combination is more general than
the equal one which can be regarded as a special case of the former.
Since the combination condition is changed, we then start with a
different form of $v_{2,q}(p_{T})$ for quarks from that in the equal
$p_{T}$ combination schemes,
\begin{eqnarray}
v_{2,q}(p_{T}) & = & 0.286p_{T}^{4},\;\;\;0\leq p_{T}\leq0.525,\label{eq:v2q-non0}\\
v_{2,q}(p_{T}) & = & 0.11\tanh[1.6(p_{T}-0.4)],\;\;\;
p_{T}>0.525.\label{eq:v2q-non}\end{eqnarray} The results for hadron
$v_{2}(p_{T})$ are shown in Fig. 4. The initial hadrons with the
same flavor contents have an identical curve. One sees in the upper
panel that the pion curve agrees with data very well. Note that Ref.
\cite{Greco:2004ex} reproduces the pion data after taking into
account the momentum distribution of hadrons, a similar strategy to
ours, and Ref. \cite{Dong:2004ve} takes into account the resonance
decay contribution to pions. In contrast, we give the results for
other hadrons also fit the data besides pions. The main difference
from the equal $p_{T}$ combination scheme is that $v_{2}(p_{T})$
spectra for hadrons at mediate $p_{T}$ split into two groups (of
mesons and baryons), i.e. there is no exact quark number scaling,
but theoretical predictions are still within the error bars. More
precise measurements are needed to distinguish the two combination
schemes in the mediate $p_{T}$ range. Obviously the fine structure
of $v_{2}(p_{T})$ at low $p_{T}$ is a joint effect of resonance
decays, the quark mass difference in $v_{2,q}(p_{T})$ and unequal
$p_{T}$ combination. The reason that the unequal $p_{T}$ combination
works well at low and mediate $p_{T}$ lies in the fact that it
splits the spectra for different hadrons at the cost of slightly
spoiling the quark number scaling. But the data are not sensitive to
such a violation of the scaling.

In Fig. \ref{cap:fig5}, we give the predictions for $\phi$ mesons
in three combination schemes. The agreement with recent STAR data
is satisfactory \cite{Cai:2005wt}. In all of our calculations, we
neglect the influence of hadron re-scattering on $v_{2}(p_{T})$,
same as in Ref. \cite{Greco:2004ex}.

\section{Summary}

We use quark combination model to study elliptic flows for directly
produced hadrons before resonant decays and the final state hadrons.
Especially we have investigated the fine structure of elliptic flows
at small $p_{T}$. We find the combination mechanism works very well
in describing elliptic flow data in low $p_{T}$ as well as in mediate
$p_{T}$ regime. In the scheme with equal $p_{T}$ combination and
equal input function for $v_{2,q}(p_{T})$ for light and strange quarks,
the data for all hadrons can be well reproduced except pions, where
the fine structure can be understood by the decay contributions. Taking
into account the mass effect of the input function $v_{2,q}(p_{T})$
for light and strange quarks and assuming unequal $p_{T}$ combination,
the overall agreement between our predictions and all available data
for hadrons including pions are satisfactory. The decay contributions
and the mass effect of the input function $v_{2,q}(p_{T})$ for quarks
play important roles. The directly produced hadrons with the same
flavor content have an identical $v_{2}(p_{T})$ spectrum, which is
an obvious feature of the quark combination mechanism.

\acknowledgments

The auhors thank Shi-yuan Li, Zuo-tang Liang, Zong-guo Si for helpful discussions. Special thanks go to Qun Wang for critically reading the manuscript with many suggestions. The work is supported in part by National Natural Science Foundation of China (NSFC) under grant 10475049.


\end{document}